\documentclass[times, twoside]{StyleBioRxiv}
\usepackage{blindtext}

\def\t{\mathrm}
\def\beq{ \begin{equation}}

\def\eeq{\end{equation}}
\def\beqar{ \begin{eqnarray} }
\def\eeqar{ \end{eqnarray} }

\def\t{\mathrm}

\setcitestyle{square}

\begin{document}

\title{Surfing vortex rings for energy-efficient propulsion}
%\shorttitle{My Template}

% Use letters for affiliations, numbers to show equal authorship (if applicable) and to indicate the corresponding author
\author[1]{Peter Gunnarson}
\author[1,2,\Letter]{John O. Dabiri}

\affil[1]{Graduate Aerospace Laboratories, California Institute of Technology, 1200 E California Blvd, 91125, Pasadena, USA}
\affil[2]{Mechanical and Civil Engineering, California Institute of Technology, 1200 E California Blvd, 91125, Pasadena, USA}

\maketitle

\begin{abstract}
Leveraging background fluid flows for propulsion has the potential to enhance the range and speed of autonomous aerial and underwater vehicles. In this work, we demonstrate experimentally a fully autonomous strategy for exploiting vortex rings for energy-efficient propulsion. First, an underwater robot used an onboard inertial measurement unit (IMU) to sense the motion induced by the passage of a vortex ring generated by a thruster in a 13,000-liter water tank. In response to the sensed acceleration, an impulsive maneuver entrained the robot into the material boundary of the vortex ring. After entrainment, the robot was propelled across the tank without expending additional energy or control effort. By advecting with the vortex ring, the robot achieved a near five-fold reduction in the energy required to traverse the tank. Using the controlled finite-time Lyapunov exponent field and corresponding Lagrangian coherent structures, we analyze and explain the initial entrainment process and the sensitivity to the starting time and position of the surfing maneuver. Additionally, linear acceleration as sensed by the onboard IMU was found to correspond with the pressure gradient of the background flow, and rotational acceleration is suggested as a method for measuring the vorticity of the vortex ring. This study serves as a proof-of-concept demonstration of the potential for onboard inertial measurements to enable efficient interaction with background fluid flows.
\end {abstract}

%\begin{keywords}
%autonomous navigation | vortex rings | energy harvesting | Lagrangian coherent structures
%\end{keywords}

\begin{corrauthor}
%\texttt{r.henriques{@}ucl.ac.uk}
jodabiri@caltech.edu
\end{corrauthor}

\section{Introduction}\label{sec:intro}

The ability to leverage background fluid flows for energy-efficient propulsion has significant implications for autonomous aerial and underwater vehicles, which play a vital role in ocean exploration \citep{wynn_autonomous_2014}, ecosystem monitoring \citep{zhang_system_2021}, drone-based inspection \citep{guerrero_uav_2012}, and many other applications. For example, rather than expending energy to swim against ocean currents, autonomous underwater vehicles can significantly improve their range and speed by maneuvering into background currents that advect the robot towards the desired destination \citep{inanc_optimal_2005,rhoads_minimum_2013}. Similarly, autonomous aerial vehicles must expend their finite onboard energy supply while flying through strong winds, which significantly limits their effectiveness \citep{watkins_ten_2020}.

If the background flow field is fully known and approximately steady, it is possible to plan efficient paths prior to deployment. Increasingly, however, autonomous vehicles are tasked with navigating in highly chaotic and unsteady flow environments, such as the gusty, separated flows around buildings in urban air environments \citep{jones_physics_2022}, or the turbulent waters underneath ice shelves for critical measurements related to climate change \citep{schmidt_heterogeneous_2023}. The unsteady nature of these flow environments necessitates real-time navigation in response to the constantly changing and chaotic background eddies. 

The high degree of unsteadiness is exacerbated by the increased utilization of small, low-cost autonomous vehicles, which have low inertia and may be slower than the dominant gusts and eddies in their surroundings. For example, aerial vehicles in urban environments can be overpowered by the vortical structures that dominate the separated flows behind buildings \citep{jones_physics_2022}. In the ocean, smaller, lower-cost vehicles are being pursued to enable more widespread coverage of the vast ocean volume (e.g., bioinspired \citep{fish_bio-inspired_2020} and biohybrid \citep{anuszczyk_electromechanical_2024} robots). Given the nascent stage of development of these systems, fluid-robot interactions at these scales of unsteadiness are not yet fully characterized and are being actively investigated (e.g., \citep{fukami_grasping_2023}). Additionally, these smaller vehicles may receive a proportionally greater benefit from intelligent navigation through background flows, given their typically reduced energy storage and propulsive abilities compared to larger platforms.

In autonomous platforms, efficient navigation must be accomplished with onboard sensing, computation, and actuation. Recent algorithmic approaches have involved data-driven techniques such as reinforcement learning, in which onboard flow measurements provided sufficient information to navigate through a variety of unsteady canonical and oceanic flow environments (e.g., \citep{bellemare_autonomous_2020,hang_interpretable_2023,gunnarson_learning_2021,masmitja_dynamic_2023,reddy_glider_2018,krishna_finite_2023}). While effective, these data-driven navigation algorithms often lack verification in physical robotic systems, particularly in ocean applications \citep{masmitja_dynamic_2023}, and in general are not guaranteed to transfer to their physical counterparts \citep{dulac-arnold_challenges_2021}. Additionally, implementing onboard flow sensors adds complexity and cost to existing vehicles. 

A promising approach for sensing background flow is to infer it indirectly from inertial measurements. In nature, it has been observed that aquatic animals such as fish use their vestibular system to detect body acceleration induced by background flows \citep{coombs_lateral_2014}. In robotic applications, inertial measurements have the advantage of being inexpensive to implement due to the ubiquity of micro-electromechanical systems (MEMS) accelerometers. For example, IMUs are often already present on quadcopters and many underwater robots for stabilization and inertial guidance. For this reason, several studies have investigated using inertial data to infer background wind flows in aerial vehicles, which was used to accomplish more accurate maneuvering \citep{oconnell_neural-fly_2022} and to detect and exploit atmospheric thermal currents \citep{reddy_glider_2018}. Additionally, amplifying sensed acceleration has been proposed as a means for efficient navigation in turbulence \citep{bollt_how_2021}. 

In this study, we use onboard inertial data to detect and exploit passing vortex rings for propulsion. We tested the strategy using the Caltech Autonomous Reinforcement Learning robot (CARL), a palm-size autonomous underwater robot \citep{gunnarson_fish-inspired_2024}. A thruster in a 13,000-liter water tank generated individual vortex rings, which served as a repeatable background flow eddy. By executing a short burst maneuver in the direction of sensed acceleration, CARL swam into the vortex ring, passively advecting with the flow structure across the tank. We estimate that the surfing strategy required one-fifth of the energy consumption compared with self-propulsion. Additionally, the controlled finite-time Lyapunov exponent field provides an explanation of the dynamics responsible for converting the small surfing maneuver into a long-distance, energy-efficient trajectory. Lastly, body rotation is found to serve as an additional indirect flow signal that can indicate background vorticity to further increase the available knowledge of the background flow. This work demonstrates a flow-based navigation strategy that closes the loop between sensing a background flow structure and exploiting that flow with an efficient and targeted maneuver.

\section{Experimental setup}

\begin{figure}[!h]
  \centerline{\includegraphics[width=0.45\textwidth]{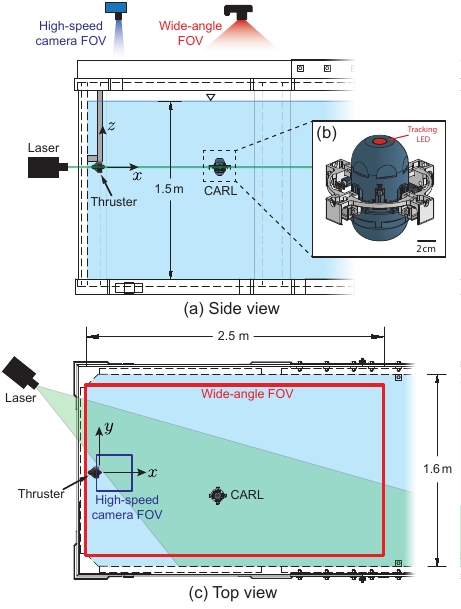}}
  \caption{Experimental setup showing a side view (\textit{a}) and top view (\textit{c}) of the water tank. A thruster mounted on a side wall of the tank ($D$ = 10$\,$cm) generates vortex rings ($\Gamma/\nu \approx$ 200,000), which are detected and exploited by CARL (\textit{b}). A wide-angle and a high-speed camera are mounted above the tank and track the position of CARL while simultaneously recording PIV measurements. A laser sheet illuminates the horizontal $x$-$y$ plane for PIV measurements.}
\label{fig:exp_setup_surfing}
\end{figure}

As a testing environment, we used a portion of a 1.5$\,$m deep, 1.8$\,$m wide, and 4.8$\,$m long water tank (Figure \ref{fig:exp_setup_surfing}). Vortex rings were generated by pulsing a thruster (Blue Robotics T200; diameter $D=10\,$cm) mounted on a wall of the tank pointing horizontally in the $x$-direction (see Figure \ref{fig:exp_setup_surfing}). The vortex rings served as a repeatable ``unit eddy'', i.e., a well-characterized vortical flow structure that could be generated on demand. To be sure, vortical flow structures in real-world flows typically involve collections of eddies of various scales, orientations, and incoming directions \citep{stutz_dimensional_2023}, but for the purposes of this study, vortex rings functioned as a repeatable flow structure for proof-of-concept demonstration and analysis. Additional details regarding the vortex ring generation and measured properties are discussed in Supplementary Material A.

To test an inertial navigation strategy in the water tank, we used CARL, a palm-sized, autonomous underwater robotic platform (Figure \ref{fig:exp_setup_surfing}b). Details of the design and construction of CARL are described in \citep{gunnarson_fish-inspired_2024} and in Supplementary Material B. CARL was equipped with an onboard inertial measurement unit (MPU-6050), which measured the linear acceleration and angular velocity of CARL. Additionally, ten motors enabled translational swimming motion in all three axes, and rotational motion about the vertical axis. At the start of each episode, CARL was manually piloted to a position approximately 3$D$ downstream and 2$D$ in the negative $y$ direction from the thruster (see Figures \ref{fig:ensemble_trajectories} and \ref{fig:LCS} for measured starting locations). CARL was then commanded to dive to the depth of the thruster and began recording onboard data. At this starting position, CARL was outside of the direct path of the vortex ring, but still close enough to sense the effects of a passing vortex ring. After CARL reached the depth of the thruster, the surfing policy described in the following section was activated, and the thruster was commanded to generate a vortex ring. After 12 seconds, CARL stopped recording data and returned to the surface, which marked the end of a trial.

\section{Vortex ring surfing strategy}

To exploit a passing vortex ring for propulsion, CARL was programmed with a simple but effective policy: if the magnitude of the acceleration in the $y$-direction exceeded a threshold, CARL would swim impulsively in the same direction as the sensed acceleration. After this maneuver, CARL typically became entrained into the vortex ring, and was propelled across the tank without requiring any additional control effort or energy expenditure. Using this ``surfing'' strategy, CARL demonstrated the ability to autonomously exploit a background flow structure for forward propulsion. 

\begin{figure*}[!h]
  \centerline{\includegraphics[width=0.85\textwidth]{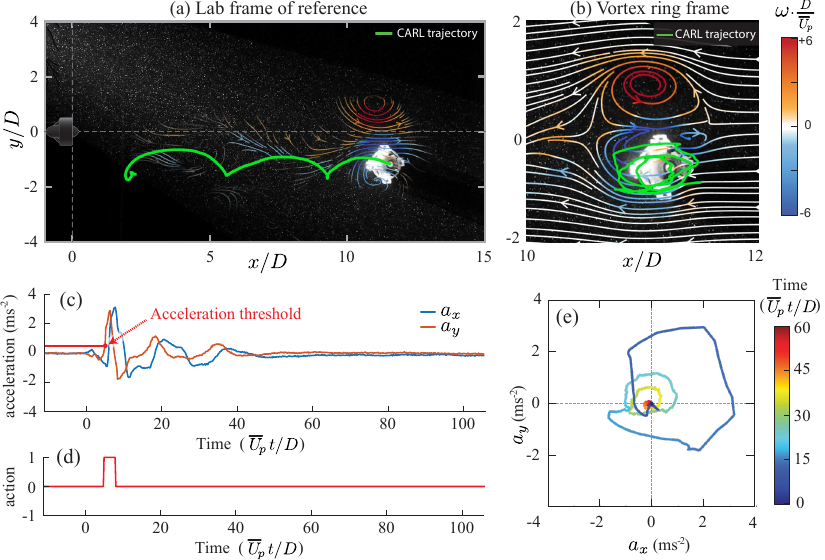}}
  \caption{Detecting and exploiting a vortex ring for propulsion. (\textit{a}) Example trajectory of CARL (green line) while surfing a vortex ring. Using simultaneous particle image velocimetry, instantaneous streamlines are plotted in the lab frame of reference and shaded by the vorticity to visualize the vortex ring. (\textit{b}) Flow streamlines are plotted in a reference frame translating with the vortex ring. The trajectory of CARL is contained inside the lower half of the vortex ring and orbits in the same direction as the local vorticity. (\textit{c}) Linear acceleration signal sensed by the IMU onboard CARL. The red line indicates the threshold used to detect the presence of a vortex ring, which autonomously triggered an impulsive maneuver (\textit{d}) with the same sign as the sensed $y$-acceleration. After the impulsive maneuver, CARL remained inside the vortex ring with no additional control effort or energy expenditure. (\textit{e}) The $x$ and $y$ components of acceleration orbit the origin, further highlighting the circular motion of CARL induced by the vortex ring. Distances are normalized by the thruster diameter ($D$ = 10 cm). Time is normalized by the vortex formation time ($D /\overline{U}_p \sim 9\,$s, \citep{gharib_universal_1998}), and $t = 0$ is defined by the origination of a vortex ring, estimated by the vortex propagation model shown in Figure \ref{fig:ensemble_trajectories}.}
\label{fig:example_trajectory}
\end{figure*}

An example trajectory and time-history of sensed acceleration are plotted in Figure \ref{fig:example_trajectory} and shown in Movies S1 and S2. As the vortex ring initially passes near CARL, it induces a body acceleration which was detected by the onboard IMU (Figure \ref{fig:example_trajectory}c). For these experiments, the orientation of CARL was fixed so that the coordinates of the reference frame onboard CARL matched the coordinate system of the tank (see Supplementary Material C for details). If the $y$-acceleration exceeded a threshold of 0.5$\,$ms$^{-2}$, CARL swam at maximum thrust for 0.3 seconds in the same direction as the sensed $y$-acceleration. This impulsive maneuver was highly successful at entraining CARL into the vortex ring. It is important to highlight that this impulsive maneuver was perpendicular to the direction of surfing motion; zero thrust contributed directly towards the forward propulsion of CARL. Additionally, the surfing strategy is fully autonomous, i.e., CARL possessed no prior knowledge of where and when the vortex ring would originate.

After successful entrainment, CARL advected passively with the vortex ring across the tank, which is visualized in Figure \ref{fig:example_trajectory}a and b using a snapshot of simultaneous particle image velocimetry (PIV, see Figure \ref{fig:exp_setup_surfing} for measurement setup). After entrainment, CARL orbited inside the lower half of the vortex ring in a clockwise direction, i.e., in the same direction as the local vorticity in that part of the vortex ring. In this sense, the translation of CARL matched that of the surrounding fluid. Throughout these experiments, CARL maintained a constant heading using feedback control, despite the presence of background vorticity.

While the navigation policy used in these experiments is simple, it demonstrates the potential of combining onboard sensing and navigation for exploiting unsteady background flows. Additionally, it is straightforward to extend this strategy beyond sensing and swimming in only the $y$-direction. For example, Figure \ref{fig:example_trajectory}e shows that both the $x$ and $y$-components of the onboard acceleration detect the motion induced by the vortex ring. When the vortex ring passes near CARL, the acceleration increases to its maximum magnitude, and afterwards orbits the origin in a clockwise direction corresponding to the direction of vorticity in the vortex ring. This circular motion can also be seen in Figure \ref{fig:example_trajectory}c, in which the phase of the $x$ and $y$ acceleration appear offset by roughly 90 degrees. In addition to linear acceleration, the potential for rotational acceleration to provide additional flow information is discussed in Section \ref{sec:body_rotation_vorticity}.

\section{Repeatability and success rate} \label{sec:repeatability}

\begin{figure}
    \centerline{\includegraphics[width=0.55\textwidth]{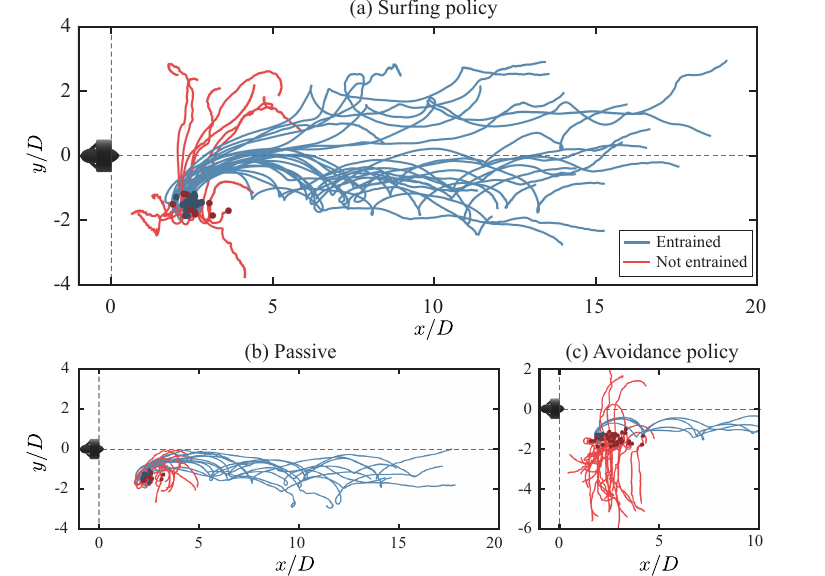}}
    \caption{Comparison of control strategies for exploiting or avoiding a passing vortex ring. (\textit{a}) Surfing policy ($N$=37). Using the surfing strategy, CARL successfully maneuvered into the vortex ring in 62\% of trials, exploiting the vortex ring for forward propulsion. (\textit{b}) For comparison, CARL was commanded to take no action ($N$=25), which results in entrainment in 48\% of episodes. (\textit{c}) CARL was commanded to avoid the vortex rings ($N$=45), resulting in a reduced entrainment rate of only 9\%.}\label{fig:ensemble_trajectories}
    \index{figures}
\end{figure}

The surfing strategy is highly repeatable, resulting in successful entrainment into the vortex ring in 62\% of trials (N=37), compared with a 48\% entrainment rate if CARL takes no action and drifts passively (N=25). Successful maneuvers are visualized in Figure \ref{fig:ensemble_trajectories} by the color of the trajectories: blue trajectories indicate that CARL was entrained into the vortex ring ($x_\mathrm{final}/D \geq 6$), and red trajectories indicate that CARL was not entrained ($x_\mathrm{final}/D < 6$). 

A successful surfing maneuver requires accomplishing the both tasks of sensing and maneuvering into the vortex ring. Approximately 19\% of the unsuccessful surfing maneuvers can be attributed to an error in sensing, in which the acceleration did not exceed the 0.5$\,$ms$^{-2}$ threshold and CARL took no action, or the acceleration pointed away from the vortex ring instead of towards it. An explanation of and potential improvements to the sensing strategy are discussed in Section \ref{sec:acceleration}. In other cases, the vortex ring was detected successfully, but the detection occurred too late for CARL to maneuver into the vortex ring. These cases are illustrated and discussed in sections \ref{sec:LCS} and Supplementary Material D. Despite these unsuccessful trials, the presence and direction of the vortex ring was correctly detected in 81\% of episodes, and the majority of surfing attempts resulted in successful entrainment into the vortex ring.

We also plot trajectories in which CARL was programmed to avoid the vortex ring by swimming in the direction opposite of the sensed acceleration (Figure \ref{fig:ensemble_trajectories}c). As with the surfing policy, the avoidance policy is highly effective: CARL avoids the vortex ring in 91\% of episodes (i.e., the entrainment rate is reduced to 9\%). 
 
The inclusion of the avoidance policy has several practical motivations. For a navigation task in the real world, flow structures may not necessarily propagate in the direction of desired motion. In that case, it would be advantageous to avoid entrainment to prevent motion in a counterproductive direction. Additionally, some aquatic animals such as jellyfish employ vortex rings for prey capture \citep{peng_transport_2009}. In turn, prey often exhibit escape maneuvers in response to deformation of the surrounding flow, and thus it may be insightful to use these robotic experiments to probe questions of predator--prey interactions. The effectiveness of the avoidance policy demonstrates that the surfing policy can be simply adjusted to test these different behaviors of exploiting or avoiding flow structures.

\section{Energy savings and momentum transfer} \label{sec:energy_savings}

Exploiting background flow structures has the potential to greatly reduce the energy required for propulsion. To estimate the propulsive energy savings of vortex ring surfing, we compare the energy required to surf a vortex ring to the energy required to travel the same distance under self-propulsion. To surf the vortex ring, CARL swims at maximum thrust for 0.3 seconds, which accelerates CARL from rest to a speed of $u_{\t{impulse}}\approx24\,$cm$\,$s$^{-1}$. Afterwards, no energy is expended to maintain position inside the vortex ring. The primary energy expenditure during this maneuver is therefore the change in kinetic energy of CARL required to accelerate CARL from rest: 

\begin{equation}
E_{\t{surf}} = \frac{1}{2} m_{\t{CARL}} (1 + \alpha_{xx}) u_{\t{impulse}}^2 \approx 16.7\,\t{mJ}, 
\end{equation}

\noindent
where $m_{\t{CARL}}$ is the mass of CARL ($\sim$355$\,$g) and $\alpha_{xx}$ is the estimated added mass coefficient of CARL in the direction of propulsion ($\sim$0.63, see Supplementary Material B for details). 

To estimate the energy required to swim the same distance without the aid of the vortex ring, we considered steady swimming at constant velocity, in which case the primary energy expenditure is overcoming steady drag while swimming at the speed $u_{\t{swim}}$. For a fair comparison, we assume that $u_{\t{swim}}$ is the same as the average speed of the vortex ring, $\overline{u}_{\t{vortex}}$, over the duration of an episode. 

\begin{equation}
    E_{\t{self-propulsion}} = \frac{1}{2} \rho \overline{u}_{\t{vortex}}^2 C_d A L\approx 82.7\,\t{mJ},
\end{equation}

\noindent
where $C_d$ is the drag coefficient ($\sim$1.1), $A$ is the frontal area of CARL ($\sim$78.3$\,$cm$^2$), and $L$ is the distance traveled during a surfing episode ($\sim$1.32$\,$m). Using these values, the energy saved by surfing the vortex ring can be estimated:

\begin{equation}
    E_{\t{self-propulsion}} / E_{\t{surf}} \approx 4.9.
\end{equation}

\noindent
By surfing the vortex ring, CARL requires approximately 4.9 times less energy than would be expended to traverse the same distance under self-propulsion.

To be sure, both cases of vortex surfing and self-propulsion could be optimized. For example, it may be possible to swim into the vortex ring with a significantly smaller impulsive maneuver depending on the initial position of CARL (see Section \ref{sec:LCS} for a discussion). Likewise, CARL has a relatively high $C_d$ of 1.1, so the energy required for self-propulsion could be reduced through streamlining. Nevertheless, our results demonstrate that, for a given vehicle design, energy can be harvested from the background flow by appropriately maneuvering in response to onboard detection of the background flow. Additionally, the thrust used for the impulsive surfing maneuver was directed in the $y$-direction, perpendicular to the forward motion of CARL after entrainment into the vortex ring. In this sense, none of the thrust generated by CARL during the impulsive maneuver directly contributed towards forward propulsion; energy for forward propulsion was derived from the surrounding flow.

\begin{figure}
    \centerline{\includegraphics[width=0.45\textwidth]{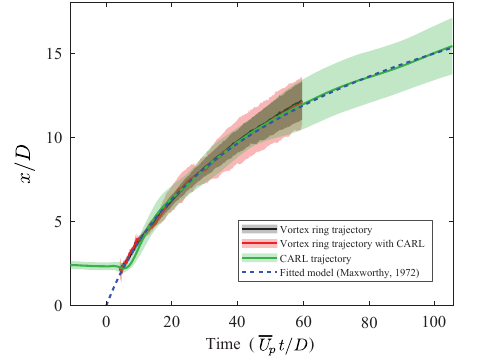}}
    \caption{Streamwise trajectories of the vortex rings and CARL. The average trajectory of the vortex ring propagating freely in the tank (black line, $N=5$) does not significantly change when CARL is entrained (red line, $N=3$), which can be explained by the relatively large mass of the vortex ring compared to CARL ($m_{\t{vortex}} / m_{\t{CARL}} \approx 22$). For comparison, the average trajectory of CARL is plotted in green ($N=35$). CARL is initially stationary, and then is entrained by a vortex ring. The Maxworthy model for vortex propagation (blue) is fitted to the average CARL trajectory, and the intersection with $x/D = 0$ is used to define the time $t = 0$. The duration of the PIV data is limited by the extent of the laser sheet, whereas CARL is visible for the entire duration of each trial.}\label{fig:vortex_trajectories}
    \index{figures}
   % \vspace{-20pt}
\end{figure}

To this point, CARL gaining energy implies a reduction in energy of the vortex ring. To see if the vortex ring is affected by this interaction, the streamwise position of vortex rings with and without the entrainment of CARL are plotted as a function of time in Figure \ref{fig:vortex_trajectories}. Within the error of our measurements, the entrainment of CARL does not significantly alter the average trajectory of the vortex rings. An example comparison is shown in Movie S3.

The apparent lack of change to the propagation of the vortex rings can be understood by considering the relative size of the vortex ring compared to CARL. With an estimated mass of $\sim$9$\,$kg and added mass coefficient of $\sim$0.44, the vortex ring is approximately 22 times as massive as CARL. As a result, the entrainment of CARL does not significantly reduce the forward momentum of the vortex ring. As another point of comparison, the total kinetic energy of the vortex ring was estimated to be approximately 1.5$\,$J (see Supplementary Information A for details), which is roughly 23 times the energy saved by CARL when surfing the vortex ring. If the vortex rings were significantly smaller, or the robot significantly larger, one might expect the entrainment of the robot to more noticeably alter the trajectory and evolution of the vortex ring.

A related inquiry is the maximum possible distance that CARL could surf until the ring loses coherence, for example, through viscous diffusion or the growth of instabilities. In Figure \ref{fig:vortex_trajectories}, a model for the trajectory of the vortex ring ($x(t) = 1/c \, \t{ln}(c u_0 \, t + 1)$, \citep{maxworthy_structure_1972}) was fitted to the CARL trajectory data, and shows good agreement with both the trajectory of CARL and the trajectories of the vortex rings for all of the available data. Additionally, simultaneous PIV measurements of CARL surfing the vortex ring show the presence of the vortex ring for the entire field of view of the wide-angle camera. Therefore, it may be possible for CARL to continue surfing the vortex rings for a longer duration than was recorded with our experimental setup. After a sufficiently long duration, the vortex rings may begin to break up or dissipate, although additional experiments are needed to determine whether the entrainment of CARL would accelerate or impede this process.

\section{Connections with Lagrangian coherent structures} \label{sec:LCS}

The primary energetic benefit of the surfing strategy arises from passive advection, i.e. CARL propagates passively with the vortex ring after entrainment. To better understand the physical mechanisms underlying this entrainment and advection process, we use the framework of Lagrangian coherent structures (LCS). LCS define material barriers in fluid flow and have been used to analyze coastal flows \citep{shadden_definition_2005}, the spread of pollution in the ocean \citep{lekien_pollution_2005}, and prey capture during jellyfish feeding \citep{peng_transport_2009}. Vortex rings can be viewed a Lagrangian coherent structure in the sense that a vortex ring forms a distinct region of transport of the fluid parcels inside the boundary of the vortex ring.

Additionally, researchers have drawn connections between LCS and optimal paths through flow fields in the context of underwater robots. For example, Inanc et al. \citep{inanc_optimal_2005} observed that energy-efficient trajectories through ocean currents in Monterey Bay coincided with the LCS of the background flow. Taking this a step further, Senatore and Ross \citep{senatore_fuel-efficient_2008} described how the LCS can be used as a tool for generating optimal energy-efficient paths through fluid flows. More recently, Krishna et al. \citep{krishna_finite_2023,krishna_finite-horizon_2022} have shown that LCS can explain key characteristics of optimal trajectories in fluid flows, such as the sensitivity of the optimal path to initial conditions and cost functions.

A common method of quantifying LCS is to compute the finite-time Lyapunov exponent (FTLE), which measures the degree to which the paths of neighboring fluid parcels diverge over a fixed time horizon. The FTLE is a scalar field whose ridges correspond to transport barriers, and therefore also the boundaries of Lagrangian coherent structures \citep{shadden_definition_2005}. The main idea of the FTLE is to cast the fluid flow as a dynamical system, in which a given fluid tracer particle with position $\mathbf{x}(t)$ moves according to the background flow:

\begin{equation}
    \dot{\mathbf{x}} = \mathbf{u}(\mathbf{x}(t),t).
\end{equation}

\begin{figure*}[!t]
    \centerline{\includegraphics[width=0.85\textwidth]{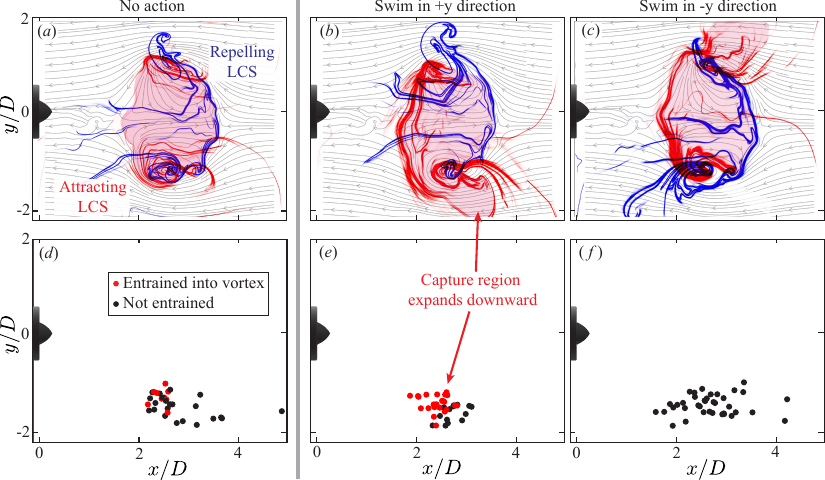}}
    \caption{LCS may explain the positional dependence of entrainment. Top row: using PIV measurements of a vortex ring, the forward time (red) and backward time (blue) FTLE field is computed for the undisturbed flow (\textit{a}) and with an impulsive maneuver towards (\textit{b}) and away from (\textit{c}) the vortex ring. Streamlines of the undisturbed flow in the vortex ring reference frame are plotted in the background. Bottom row: starting locations of CARL. Entrained points (red) correspond qualitatively with the attracting LCS regions.}\label{fig:LCS}
    \index{figures}
\end{figure*}

\noindent
For a given point in the fluid domain, the flow map, $\Phi_0^T$, is defined as the mapping between the starting position of a particle to the position after being advected by the background flow for a finite time $T$:

\begin{equation}
    \Phi_0^T : \mathbf{x}(0) \mapsto \mathbf{x}(0) + \int_0^T \mathbf{u}(\mathbf{x}(t),t)\t{d}t.
\end{equation}

\noindent
The FTLE is then computed by taking the largest singular value of the Jacobian of the flow map, which we estimate using finite differences on a grid of virtual particles advected according to the measured flow field: 

\begin{equation}
    \sigma(\mathbf{x};\Phi_0^T) = \frac{1}{|T|} \t{log} \sqrt{\lambda_{\t{max}}(\mathbf{D}\Phi_0^T)^{\ast}(\mathbf{D}\Phi_0^T)}.
\end{equation}

\noindent
LCS can be then extracted by computing ridges of the FTLE field. Here, we simply threshold the value of the FTLE field to visualize coherent structures. 

The concept of the FTLE field can be extended to incorporate the active motion of swimmers or the inertia of particles in the flow. For example, this has been referred to as the control FTLE (cFTLE, e.g., \citep{krishna_finite-horizon_2022}) or particle LCS (pLCS, e.g., \citep{peng_transport_2009}), where each fluid tracer is modeled as moving according to the background flow and the swimmer's self-propulsion. In our case: 

\begin{equation} \label{eq:FTLE_dynamics}
    \dot{\mathbf{x}} = \mathbf{u}(\mathbf{x}(t),t) + \mathbf{u}_{\t{CARL}}(t),
\end{equation}

\noindent
where $\mathbf{u}_{\t{CARL}}(t)$ is the motion due to CARL executing the impulsive surfing maneuver. The effects of the inertia of CARL could also be included in this equation for greater quantitative accuracy (e.g., \citep{peng_transport_2009}), but the model in Equation \ref{eq:FTLE_dynamics} nevertheless offers qualitative insight into dynamics of this vortex surfing strategy. The cFTLE effectively shows transport boundaries for the swimmer rather than just the surrounding fluid, which gives insight into the swimmer trajectories and their sensitivity to initial conditions. In the above formulae, particles are integrated forward in time, which results in repelling LCS, or boundaries over which particle paths diverge. Integrating particles backwards in time computes structures that are attracting, which help indicate particle entrainment into the vortex rings.

In Figure \ref{fig:LCS}a, we plot snapshots of the attracting and repelling LCS of the vortex ring from PIV measurements. As expected, the LCS delineates the material boundary of the vortex ring. Below in panel (d), we plot starting locations in which CARL took no action and drifted passively in the flow (i.e., $\dot{\mathbf{x}} \approx \mathbf{u}$). The starting positions are shaded red if CARL was entrained into the vortex ring.

To analyze the surfing maneuver, we plot the cFTLE for CARL swimming impulsively into the vortex ring (Figure \ref{fig:LCS}b) and away from the ring (Figure \ref{fig:LCS}c). In these cases, the velocity of the swimmer $\mathbf{u}_{\t{CARL}}$ is directly copied from position data of CARL executing the impulsive surfing maneuver in the appropriate direction. When swimming into the vortex ring, the attracting region extends downwards in the negative $y$ direction. Correspondingly, the percentage of starting points that are captured by the vortex ring increases from 48\% with passive swimming to 73\% with a maneuver towards the vortex ring (Figure \ref{fig:LCS}e). Conversely, when CARL executes a maneuver to avoid the vortex ring (Figures \ref{fig:LCS}b and \ref{fig:LCS}c), the attracting cLCS shifts upwards, while the repelling cLCS moves downwards, and none of the starting locations are entrained.

The starting positions for the vortex surfing case appear spatially clustered, with entrained (red) points positioned closer to the centerline. The cLCS appears to explain this division: points lying within the attracting cLCS are entrained, while points outside are not entrained. In this way, the cLCS may explain and predict the entrainment of CARL into the vortex ring for a given initial position and maneuver. Additionally, by simply altering $\mathbf{u}_{\t{CARL}}(t)$ in Equation \ref{eq:FTLE_dynamics}, the effectiveness of an arbitrary surfing maneuver could be estimated and optimized prior to deployment. 

The FTLE field also explains why a small impulsive maneuver from CARL can have a large impact on final position and energy savings. Ridges of the FTLE field correspond to regions of high flow stretching, in which neighboring particle paths diverge exponentially \citep{shadden_definition_2005}. Therefore, a small jump in position, such as the impulsive surfing maneuver, can be amplified into a large change in the trajectory when executed near the boundary of the vortex ring. Additionally, Krishna et al. \citep{krishna_finite-horizon_2022} observed that for energy-efficient trajectories, there are spikes in control effort and cost function that correlate with LCS boundaries because of this amplification of small perturbations at LCS boundaries. Our experiments demonstrate this principle in practice: a small spike in thrust near the LCS boundary of the vortex ring results in large energy savings for forward propulsion.

LCS help explain the sensitivity of trajectories to their initial conditions, such as the starting position of CARL. However, in unsteady flow fields, such as a passing vortex ring, trajectories near boundaries of the LCS are also highly sensitive to the starting time \citep{lekien_pollution_2005}. In other words, surfing a vortex ring requires being in the right place at the right time. To illustrate the combined spatial and temporal dependence, several example CARL trajectories are plotted in the Supplementary Information D and Movie S4.

\section{Analysis of acceleration as a sensory input} \label{sec:acceleration}
As demonstrated in the previous section, the final trajectory of CARL depends on the timing and location of the surfing maneuver, which is triggered by the $y$-component of the acceleration measured by the IMU onboard CARL (see Figure \ref{fig:example_trajectory}). Therefore, in this section, we seek to model and understand how acceleration is used as a signal to detect the presence and location of a passing vortex ring. 

\begin{figure*}[!h]
    \centerline{\includegraphics[width=0.7\textwidth]{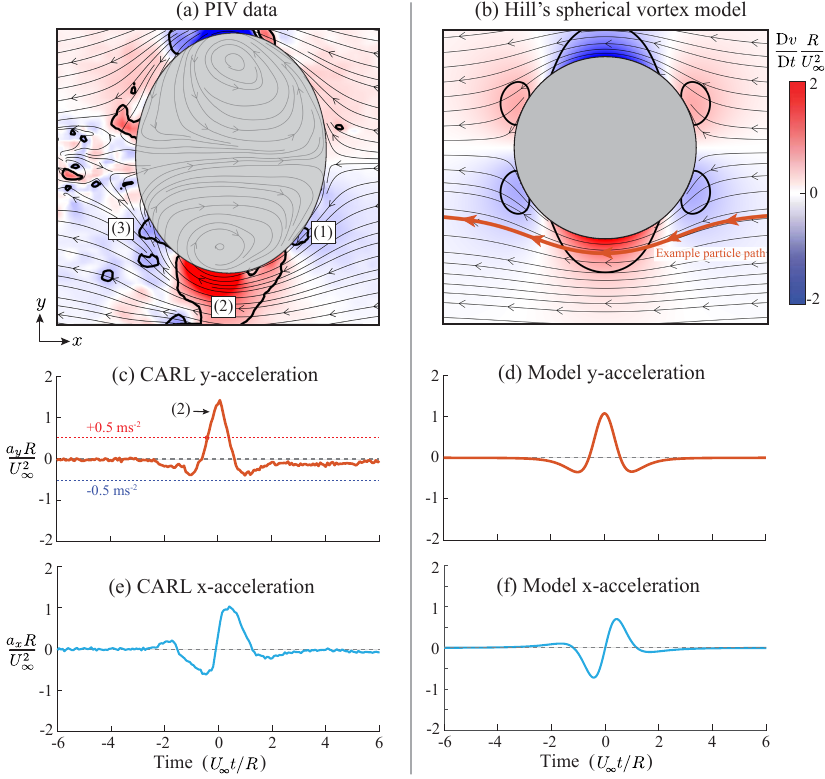}}
    \caption{Comparison between the measured and modeled acceleration signal from a passing vortex ring. (\textit{a}) The $y$-component of the material derivative is plotted using PIV data of a vortex ring, indicating the acceleration experienced by ideal fluid tracers. For example, particles in regions (1) and (3) experience negative $y$-acceleration, while particles in region (2) experience larger and positive $y$-acceleration, potentially explaining the signals sensed by CARL. Contours of +0.5 and -0.5$\,$ms$^{-2}$ are also plotted, which correspond to threshold used by CARL to execute the surfing maneuver. An ellipse is overlaid to indicate the approximate boundary of the vortex ring. (\textit{c}) Acceleration in the $y$-direction sensed by CARL as a vortex ring passes nearby. The spike in positive $y$-acceleration may correspond to region (2) in the panel above, and is used to locate the vortex ring during a surfing maneuver. (\textit{b}) Hill's spherical vortex model. The regions of positive and negative acceleration agree with the PIV data qualitatively. Both components of the acceleration of a ideal tracer particle traveling through the potential flow model (\textit{d},\textit{f}) qualitatively agree with the signal sensed by the IMU onboard CARL (\textit{c},\textit{e}). Acceleration values are normalized by $U^2_{\infty}/R \approx 1.05\,$ms$^{-2}$, where $U_{\infty}$ is the vortex ring propagation speed and $R$ is the minimum radius of curvature of the vortex ring boundary. Time $t=0$ is defined by the peak in $y$-acceleration in (c) and (d).}\label{fig:acceleration}
\end{figure*}

In Figure \ref{fig:acceleration}a, we plot the $y$-component of the material derivative, $\t{D} v/\t{D}{t}$, computed from PIV measurements, which represents the $y$-component of acceleration experienced by ideal tracer particles in the flow. While CARL is by no means an ideal tracer particle due to having finite size and inertia, the acceleration of the background flow provides a model of the acceleration of CARL at any spatial location in the PIV domain. An ellipse is overlaid to indicate the approximate boundary of the vortex ring, since this section considers only the signals initially sensed by CARL before deciding to maneuver, i.e., before entrainment. Streamlines in the vortex ring reference frame are also plotted to show the direction of flow. Additionally, contours of +0.5 ms$^{-2}$ and -0.5 ms$^{-2}$ of $\t{D} v/\t{D}{t}$ are included for direct comparison with the acceleration threshold used to trigger the surfing maneuver. 

For comparison, the $y$-acceleration recorded by CARL is plotted below in Figure \ref{fig:acceleration}c during an episode in which CARL passively drifted as a vortex ring passed nearby (e.g., Figure \ref{fig:ensemble_trajectories}b). The recorded acceleration exhibits features that correspond to regions of the PIV data. As the vortex ring initially approaches, CARL records a small, negative $y$-acceleration. This corresponds to region (1) in the PIV data, in which streamlines bend away from the front of the vortex ring. Next, CARL experiences a large positive $y$-acceleration, which corresponds with region (2) in the flow, as the trajectory of CARL curves around the vortex ring boundary. Finally, the acceleration becomes negative again, which corresponds with region (3). It may be that the acceleration sensed by CARL corresponds the curvature of the streamlines bending around the vortex ring boundary. 

To model the acceleration of CARL analytically, we consider Hill’s spherical vortex model \citep{hill_spherical_1894} and plot the resulting flow acceleration in Figure \ref{fig:acceleration}b. The regions of positive and negative acceleration in the vortex ring model match the PIV data qualitatively. For example, the contour of 0.5$\,$ms$^{-2}$ acceleration encompassing region (2) in panel (a) also appears in the modeled vortex ring. While the shape of the physical vortex ring is ellipsoidal rather than spherical, the overall trends of the streamlines curving around the boundary of the vortex ring are the same.

In panels (d) and (f), the $x$ and $y$-acceleration are plotted for an ideal tracer particle moving through the modeled flow. Despite being an idealized model, the particle acceleration in (d) and (f) qualitatively matches all features of the acceleration recorded by CARL in (c) and (e). To summarize, the acceleration experienced by CARL can be modeled by considering the flow acceleration around the boundary of the vortex ring.

With this modeling framework, the vortex ring detection rates discussed in section \ref{sec:repeatability} can now be understood. CARL correctly identified the location of the vortex rings in 81\% of trials, i.e., the acceleration exceeded the 0.5 m$\,$s$^{-2}$ threshold and also pointed towards the vortex ring. This corresponds to sensing the large acceleration in region (2) of Figure \ref{fig:acceleration}a. In this region, the $y$ acceleration points towards the vortex ring, and therefore the surfing maneuver is also directed towards and into the vortex ring. 

In 5\% of episodes, CARL swam away from the vortex ring, due to sensing the negative $y$ acceleration in region (1). The smaller size of this negative $y$-acceleration region explains why correct detections were more likely: with random variations in the starting position, CARL will encounter the larger region (2) more often than the smaller region (1). Additionally, in 2\% of episodes, the $y$-acceleration did not exceed the threshold value, and CARL did not execute a surfing maneuver. These cases may correspond to being outside of either region (1) or (2), i.e. being too far from the vortex ring to detect a strong enough acceleration signal. In this way, the contours of acceleration may explain and predict the dependence of the sensed acceleration on the position of CARL relative to the vortex ring. 

Since the initial detection of the vortex ring depends on the flow outside of the vortex ring boundary, the particular vortex ring model may not be of great importance. For a vortex ring propagating in quiescent flow, the flow outside of the vortex ring is free of vorticity, with the exception of a small trailing wake. Thus, the flow upstream and outside of the vortex ring can be modeled as potential flow, and indeed, the flow outside Hill’s spherical vortex ring is identical to the potential flow past a sphere. As a consequence, the qualitative trends in acceleration and the curvature of streamlines may be similar across various vortex ring models, or for that matter, the potential flow around a round, translating body. 

Furthermore, acceleration sensing in this context can considered a form of pressure sensing. In the potential flow outside of the vortex ring boundary, the flow acceleration is balanced by the pressure gradient:

\begin{equation} \label{eq:potential_flow}
    \frac{\t{D}\mathbf{u}}{\t{D}t} = -\frac{1}{\rho} \nabla P.
\end{equation}

\noindent
In other words, the acceleration experienced by an idealized tracer particle exterior to the vortex ring is directly proportional to the pressure gradient in the background flow. For an object with finite size such as CARL and in the absence of viscous forces, body acceleration can be attributed to the pressure field integrated around the body of the object. In this sense, acceleration sensing has a direct correspondence with pressure sensing. 

An advantage of modeling the flow exterior to the vortex ring with potential flow is that the pressure gradient, and therefore acceleration field, can be computed analytically. Scaling laws can then be developed in order to better generalize the surfing strategy, for example, by determining the required sensitivity of the onboard accelerometer for a vortex ring of arbitrary size, speed, and relative distance. To demonstrate, we compute the pressure field exterior to Hill's spherical vortex using the Bernoulli equation applied to the velocity field: 

\small
\begin{equation}
    \frac{u_r}{U_{\infty}} = \left[1-\left(\frac{R}{r}\right)^3\right] \cos(\theta), \quad  \frac{u_{\theta}}{U_{\infty}} = -\left[1+\frac{1}{2}\left(\frac{R}{r}\right)^3\right] \sin(\theta)
\end{equation}

\begin{equation} \label{eq:hill_pressure}
\begin{split}
    \frac{P - P_{\infty}}{\frac{1}{2} \rho U_{\infty}^2} = 1 - \frac{|\textbf{u}|^2}{U_{\infty}^2} = -&\left(\frac{R}{r}\right)^3\left(1-3\cos^2(\theta)\right)  \\ -& \frac{1}{4}\left(\frac{R}{r}\right)^6\left(1+3\cos^2(\theta)\right),
\end{split}
\end{equation}
\normalsize

\noindent
where $P$ is the static pressure, $P_{\infty}$ is free-stream pressure, $U_{\infty}$ is the propagation speed of the spherical vortex, $R$ is the radius of the spherical vortex, $r$ is the distance from the center of the vortex ring (here, $r \geq R$), $\theta$ is the polar angle in spherical coordinates, and $u_r$ and $u_{\theta}$ are the radial and polar components of velocity, respectively. From this pressure field, two quantities of interested can be computed.

First, we compute the maximum flow acceleration, which predicts the largest acceleration signal that could be detected by the robot within this idealized model. The maximum flow acceleration occurs on the sides of the vortex ring ($r = R, \theta = \pm90^{\circ}$), which corresponds to the region of high acceleration in region (2) in Figure \ref{fig:acceleration}. Taking the gradient of Equation \ref{eq:hill_pressure}:

\begin{equation}
    \frac{1}{\rho} |\nabla P|_{\mathrm{max}} = \left|\frac{\mathrm{D}\textbf{u}}{\mathrm{D}t}\right|_{\mathrm{max}} = \frac{9}{4} \frac{U_{\infty}^2}{R}.
\end{equation}

\noindent
For a vortex ring with a given radius $R$ and propagation speed $U_{\infty}$, this equation provides a prediction of the maximum flow acceleration. Intuitively, this equation takes the form of a centripetal acceleration, since the flow acceleration around the vortex ring is related to the streamline curvature. For this reason, the acceleration in Figure \ref{fig:acceleration} is normalized by $U^2_{\infty}/R$. For the physical vortex ring, which is not spherical, $R$ is chosen to be the radius of minimum curvature of the bounding ellipsoid, since this radius corresponds to the region of maximum acceleration in region (2) of Figure \ref{fig:acceleration}.

The maximum acceleration can also be computed in terms of the circulation of the vortex ring. For Hill's spherical vortex model, the circulation is related to the propagation speed of the vortex ring by $\Gamma = 5 U_{\infty} R$. Therefore, the maximum acceleration can be expressed as:

\begin{equation}
\left|\frac{\mathrm{D}\textbf{u}}{\mathrm{D}t}\right|_{\mathrm{max}} = \frac{9}{100} \frac{\Gamma^2}{R^3}.
\end{equation}

\noindent
These expressions for the maximum acceleration could be useful for designing the sensing capabilities of robots. For example, a vortex ring that is too large or too weak may not be detectable for a given sensitivity to acceleration.

Another useful scaling to consider is the dependence of flow acceleration on distance to the vortex ring. For example, if a robot is too far from a vortex ring, the acceleration signal may be too weak to detect with onboard accelerometers. To investigate this scaling, we consider the pressure gradient when the robot is far from the vortex ring ($r \gg R$):

\begin{equation}
    \frac{1}{\rho}\left|\nabla P\right| = \left|\frac{\mathrm{D}\textbf{u}}{\mathrm{D}t}\right| \approx \frac{3}{2} \frac{U_{\infty}^2}{R} \left(\frac{R}{r}\right)^4 \sqrt{1 - 2\cos^2(\theta) + 5\cos^4(\theta)}.
\end{equation}

\noindent
Therefore, the magnitude of the flow acceleration scales with distance to the vortex ring according to:

\begin{equation}
    \left|\frac{\mathrm{D}\textbf{u}}{\mathrm{D}t}\right| / \left|\frac{\mathrm{D}\textbf{u}}{\mathrm{D}t}\right|_{\mathrm{max}} \sim  \left(\frac{R}{r}\right)^4 , \quad (r \gg R).
\end{equation}

\noindent
In other words, the acceleration signal decays proportional to $r^{-4}$. The required sensitivity of an onboard accelerometer therefore increases correspondingly with distance from the vortex ring. 

Interpreting body acceleration as a form of pressure sensing requires several caveats. First, the inertia of CARL results in a lag between the motion of CARL and the background flow, effectively low pass filtering temporal variations in the pressure. To address this limitation, the robot’s acceleration could be estimated using models of small particles with inertia (e.g., \citep{maxey_equation_1983}) or by empirically modeling the inertia and drag forces on CARL. The inclusion of inertia may be particularly necessary in aerial applications, in which vehicles are typically significantly denser than the surrounding air. Second, the acceleration of CARL represents the pressure integrated over the body of CARL, effectively averaging out any pressure fluctuations of a scale significantly less than the size of CARL. However, even with these caveats, the results in Figure \ref{fig:acceleration} suggest that a potential flow model and ideal tracer particles capture the dominant mechanisms behind the acceleration signal for these experiments.

% To account for the finite size of CARL, the pressure field could be integrated over a circular or spherical body that approximates the shape of CARL.

\section{Body rotation and vorticity sensing} \label{sec:body_rotation_vorticity}

In addition to linear acceleration, another sensor input that could provide additional flow information is rotational acceleration due to vorticity in the background flow, sensed via the gyroscopes that are often included alongside accelerometers in low-cost IMUs. For example, Reddy et al. \citep{reddy_glider_2018} used the flow-induced rolling moment of a glider to sense shear caused by thermal plumes in the atmosphere. In our experiments, CARL prevented body rotation about the vertical ($z$) axis using a PID feedback loop, which took the angular velocity from the IMU as an input and outputted a rotational control signal to the propellers. The magnitude and direction of this rotational control signal, which we denote as $\tau_{\t{control}}$, was therefore correlated to the torque applied to CARL by the surrounding fluid, potentially enabling an indirect measurement of background flow vorticity.

\begin{figure}[!h]
    \centerline{\includegraphics[width=0.5\textwidth]{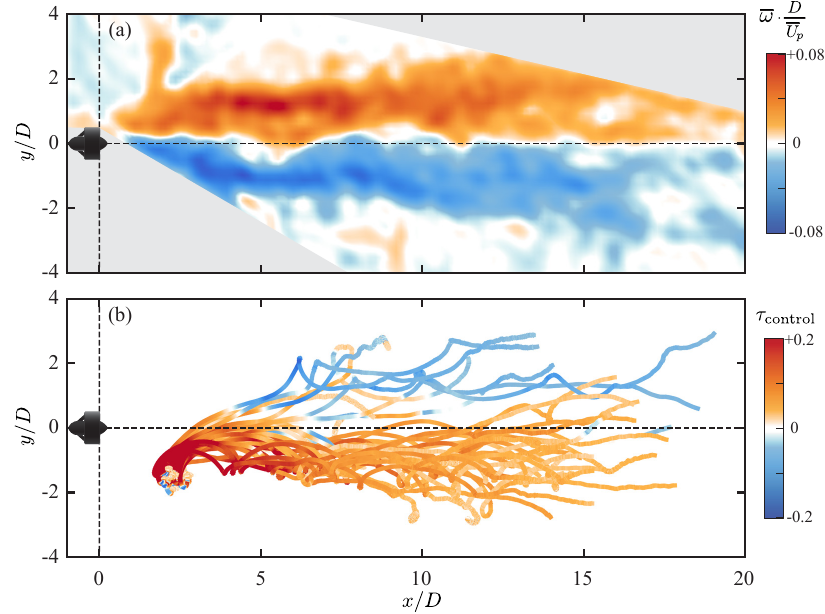}}
    \caption{Correspondence between the time-averaged vorticity and the rotational control applied to prevent CARL from rotating. (\textit{a}) Time-averaged vorticity field computed from eight consecutive vortex rings without the presence of CARL. (\textit{b}) Trajectories in which CARL surfed the vortex ring, shaded by the rotational control signal ($\tau_{\t{control}}$). The sign of the rotational control signal opposes that of the background vorticity, suggesting that body rotation could be used as an additional signal for detecting background flow structures.} \label{fig:mean_vorticity}
    \index{figures}
\end{figure}

To test this idea, we plot the time-averaged vorticity ($\overline{\omega} = 1/T \int_0^T \omega \t{d} t$) computed from PIV data of eight consecutive vortex rings in Figure \ref{fig:mean_vorticity}a. Because the vortex rings propagate across the tank in the $x$ direction, the time-averaged vorticity is positive (counter-clockwise) above the $y$-axis, and negative (clockwise) below the $y$-axis. For comparison, trajectories from 109 episodes of CARL surfing vortex rings are plotted in Figure \ref{fig:mean_vorticity}b, with the trajectories shaded by the rotational control signal commanded by CARL. 

The sign of the rotational control signal appears to oppose the background vorticity, typically being positive (counter clockwise) below the $y$-axis, and negative (clockwise) above the $y$-axis. This suggests that rotational acceleration, or rotational control effort, could be used as a supplemental inertial flow measurement technique to detect the presence of the vortex ring and the direction of the background vorticity. An accurate estimation of the background vorticity from $\tau_{\t{control}}$ would require modeling the corresponding dynamics and transfer function, but as a proof-of-concept, a quantitative comparison using a linear scaling is shown in Supplementary Material E.

\section{Discussion} \label{sec:conclusion}

In this study, we implemented energy-efficient flow-based navigation in a physical tank and robot. Using an onboard accelerometer, our robot successfully identified and maneuvered onto vortex rings, exploiting the background fluid flow for a near fivefold reduction in energy required for forward propulsion. These results demonstrate the potential of accomplishing flow sensing using the inexpensive accelerometers found in countless drones, phones, and other consumer devices. Additionally, simultaneous PIV measurements of the background flow and tools such as LCS give insight into the fluid mechanic principles underlying the surfing strategy and entrainment into the vortex ring. 

Both tasks of sensing and maneuvering could be explored further in several directions. First, an immediate extension of the current experimental setup would be to utilize both components of acceleration as well as rotation to detect the vortex ring with greater reliability. It may also be possible to complement onboard inertial sensing with direct flow measurements from flow sensors (e.g. pressure sensors) to resolve the higher-frequency temporal and spatial scales of the background flow that are filtered by the inertia and finite size of the robot. For example, studies have investigated the techniques for using arrays of pressure sensors to locate oscillating spheres \citep{dehnhardt_seal_1998,dagamseh_imaging_2013} and identify locations of vortex cores \citep{eldredge_bayesian_2024}. 

In addition to exploring the sensing problem, the surfing maneuver could be optimized using path planning algorithms (e.g., \citep{lolla_time-optimal_2014}) or learned in situ with data-driven methods such as reinforcement learning. Combining these tools of optimal maneuvering with the analytical model of the sensed acceleration could fully close the loop between flow-based sensing and navigation. Beyond optimizing the vortex ring surfing maneuver, it is useful to optimize point-to-point navigation across the entire flow environment. For example, if CARL were tasked with navigating to a particular location rather than just downstream, it may be optimal to surf the vortex ring for a short period of time, or even avoid it entirely, depending on the direction of the incoming vortex ring relative to the location of the target. Such optimizations could be performed over the Pareto front that defines tradeoffs between energy consumption and minimal travel time for point-to-point navigation \citep{krishna_finite-horizon_2022}.

It is important to generalize our results to other types of flows for implementation in real-world applications. Fortunately, the FTLE analysis in this study is not specific to any particular flow field and could be used as a trajectory prediction tool for other flows and maneuvers. Similarly, the pressure gradient could provide a map of the available acceleration signal for a general flow field. To extend beyond an isolated vortex ring, it would be informative to test navigation strategies in canonical vortical flows, such as a von K\'{a}rm\'{a}n vortex street, a turbulent wake, or double gyre flow. Of particular practical application are highly chaotic and turbulent flows, for example, near ice shelves in the ocean or in urban air environments, which involve many vortices of varying strengths, sizes, and orientations \citep{stutz_dimensional_2023}. Dimensionality reduction techniques may be key for collapsing the large space of possible incoming flow structures down to a few key parameters \citep{fukami_grasping_2023}. Alternatively, identifying and exploiting the nearest vortical structure (e.g, \citep{eldredge_bayesian_2024}) may provide sufficient information to optimize over short time horizons and plan energy-efficient paths. The results here serve as a proof-of-concept demonstration with a physical robot and prototypical flow structure, which could be extended in future work.

\section*{Supplementary Material}
Supplementary material will be available upon publication.

\section*{Funding}
This work was supported by the National Science Foundation Alan T. Waterman Award and NSF Graduate Research Fellowship Grant No. DGE 1745301.

\section*{Author contributions statement}
P.G. and J.O.D. conceived of project, P.G. conducted experiments, P.G. and J.O.D. analyzed results and wrote the paper.

%\section{Preprints}
%A preprint of this article is published at [DOI].

\section*{Data availability}
All data generated and discussed in this study are available within the article and its supplementary files, or are available from the authors upon request.

\section*{References}

\renewcommand*{\bibfont}{\small}

\end{document}